\documentclass[prl,longbibliography, twocolumn, showpacs, preprintnumbers,amsmath,amssymb,superscriptaddress,showkeys]{revtex4-1}
\usepackage{graphicx}
\usepackage{dcolumn}
\usepackage{bm}
\usepackage{subfigure}
\usepackage{multirow}
\usepackage{amsmath}
\usepackage{mathrsfs}
\usepackage{chemfig}
\usepackage{mathtools}
\usepackage{relsize}
\usepackage{color}
\usepackage{xcolor}
\usepackage{soul}
\usepackage[normalem]{ulem}
\newcommand{\rsmath}[1]{\bgroup\markoverwith{\textcolor{red}{\rule[0.5ex]{2pt}{0.4pt}}}\ULon {\textcolor{red}{#1}}}                                           

\usepackage{eso-pic}
\AddToShipoutPictureBG*{%
  \AtPageUpperLeft{%
    \hspace{\paperwidth}%
    \raisebox{-\baselineskip}{%
}}}%

\allowdisplaybreaks[4]

\begin{document}

\title{The role of receptor uniformity in multivalent binding}

\author{Xiuyang Xia}
\affiliation{School of Chemistry, Chemical Engineering and Biotechnology, Nanyang Technological University, 62 Nanyang Drive, Singapore 637459}
\affiliation{Division of Physics and Applied Physics, School of Physical and Mathematical Sciences,Nanyang Technological University, 21 Nanyang Link, Singapore 637371}%

\author{Ge Zhang}
\email{gzhang37@cityu.edu.hk}
\affiliation{Department of Physics, City University of Hong Kong, Hong Kong, China}

\author{Massimo Pica Ciamarra}
\affiliation{Division of Physics and Applied Physics, School of Physical and Mathematical Sciences,Nanyang Technological University, 21 Nanyang Link, Singapore 637371}

\author{Yang Jiao}
\affiliation{Materials Science and Engineering, Arizona State University, Tempe, AZ 85287, USA}

\author{Ran Ni}
\email{r.ni@ntu.edu.sg}
\affiliation{School of Chemistry, Chemical Engineering and Biotechnology, Nanyang Technological University, 62 Nanyang Drive, Singapore 637459}

             
\begin{abstract}
Multivalency is prevalent in various biological systems and applications due to the superselectivity that arises from the cooperativity of multivalent binding. Traditionally, it was thought that weaker individual binding would improve the selectivity in multivalent targeting. Here using analytical mean field theory and Monte Carlo simulations, we discover that for receptors that are highly uniformly distributed, the highest selectivity occurs at an intermediate binding energy and can be significantly greater than the weak binding limit. This is caused by an exponential relationship between the bound fraction and receptor concentration, which is influenced by both the strength and combinatorial entropy of binding. Our findings not only provide new guidelines for the rational design of biosensors using multivalent nano-particles but also introduce a new perspective in understanding biological processes involving multivalency.

\end{abstract}


\maketitle

\section{Introduction}
Multivalent interactions play a crucial role in a variety of biological processes~\cite{mammen1998polyvalent,boudreau1999extracellular,huskens2006multivalent,fasting2012multivalency,satav2015effects,karimi2018integrin}. 
They provide an  ``on-off" binding at a threshold receptor density, creating a biological barcode - targeting surfaces that have a receptor density above the threshold while leaving others untouched.
As a result, the multivalent binding strategy is also widely used in many biorelated applications, particularly in drug delivery~\cite{baker2010homing,bartlett2007impact,davis2010evidence,akhtar2014targeted,koenig2021structure} and bio-sensing~\cite{zeng2012carbohydrate,zhou2014multivalent,cai2014engineering}. 

The Martinez-Veracoecha and Frenkel (MF) model provides a selectivity parameter $\alpha = {\rm d} \ln \theta /{\rm d} \ln n_R$ quantifying the dependence of targeted adsorption $\theta$ on the receptor density $n_R$~\cite{martinez2011designing}. Generally, the maximum of the selectivity parameter $\alpha_{\rm max}$, where the targeted adsorption grows fastest, is defined as the selectivity employed to characterize the overall selectivity indicating the onset of guest nanoparticle binding and clustering ~\cite{martinez2011designing,albertazzi2013spatiotemporal,angioletti2017exploiting}. 
If the selectivity $\alpha_{\max} > 1$, the binding of nano-particles is superselective, which is a signature of multivalent binding.
The MF model predicts that $\alpha_{\rm max}$ increases as the binding strength becomes weaker when neglecting nonspecific interactions~\cite{martinez2011designing,curk2017optimal}, which was also observed in recent experimental systems including DNA coated colloids~\cite{scheepers2020multivalent,linne2021direct,chaikin2018},  multivalent guest-host polymers~\cite{albertazzi2013spatiotemporal,dubacheva2014superselective,dubacheva2015designing,dubacheva2019multivalent} and influenza virus particles~\cite{overeem2020hierarchical}. 

All studies mentioned above assume that the receptors grafted on the host substrate follow the Poisson distribution considering they are spatially uncorrelated. However, due to the complex environment on cell membranes, the receptors are heterogeneously distributed and correlated~\cite{lingwood2010lipid}, of which the effect remains unknown. {Additionally, recent breakthroughs in DNA nanotechnol- ogy offer the possibility to precisely design the spatial distribution of receptors on a substrate~\cite{patjacs}.}
Here we investigate how the uniformity of receptor distribution affects the selectivity in multivalent nano-particle binding by focusing on the hyperuniform, Poisson, and anti-hyperuniform distributions~\cite{torquato2021local}. We find that the more uniformly distributed receptors lead to higher selectivity $\alpha_{\max}$, and intriguingly, the maximum selectivity appears at certain intermediate binding energy for hyperuniform distributions, which is qualitatively different from the Poisson distribution and anti-hyperuniform distributions with $\alpha_{\max}$ approaching the upper bound at the infinitely weak binding energy limit. Moreover, the highest selectivity obtained for receptors of hyperuniform distributions can be significantly larger than the upper bound in the Poisson and anti-hyperuniform distributions, where the relatively large number fluctuation of receptors masks the effect and causes the selectivity to increase monotonically with decreasing the binding strength.

\section*{Methodology}
\begin{figure}[!t]
\centering
        \includegraphics[width=0.45\textwidth]{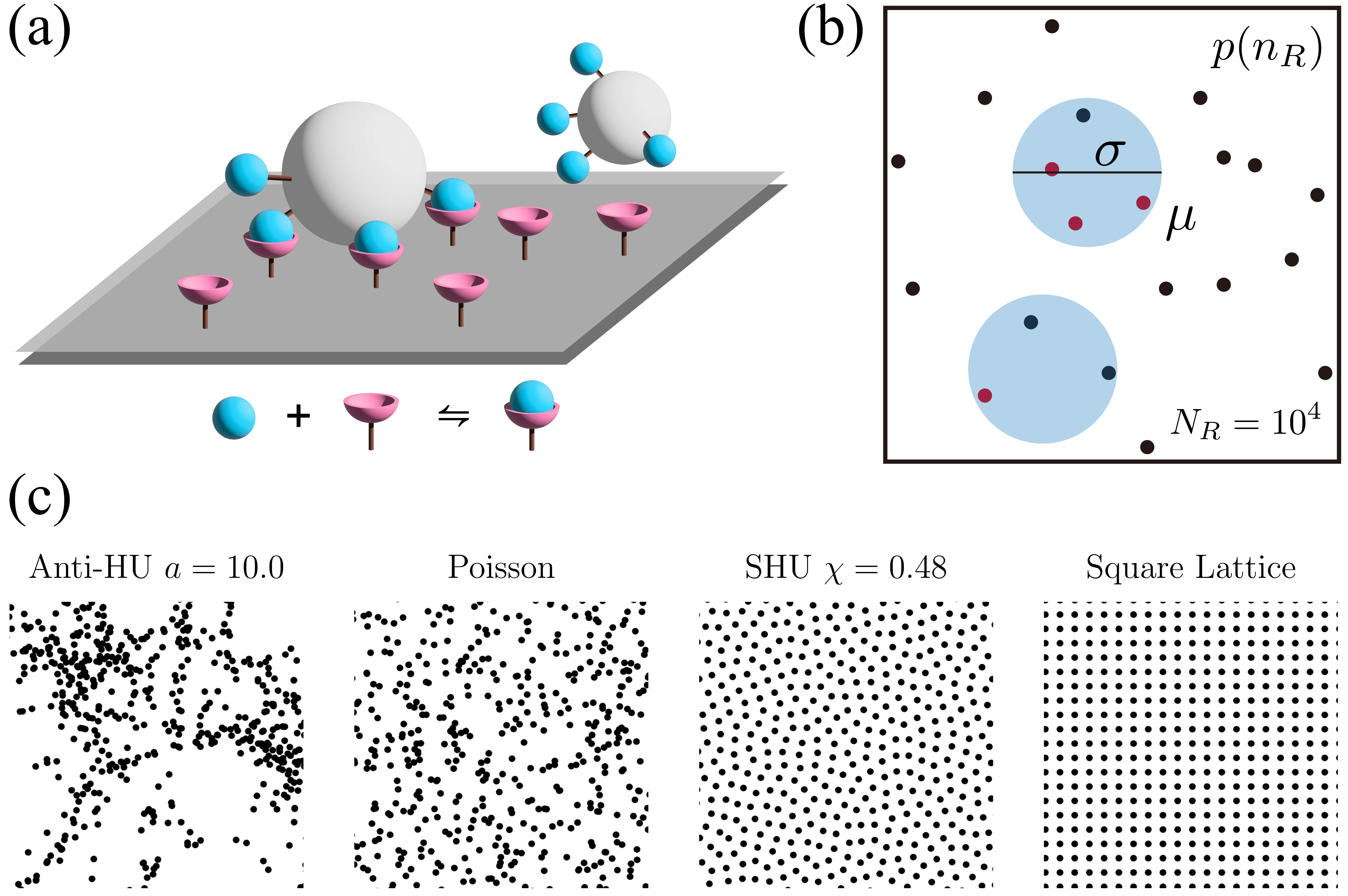}
    \caption{\label{Fig1} \textbf{Multivalent nano-particle binding.} (a) Schematic representation of the prototypical multivalent adsorption model, in which the ligands (blue) on the particles (white) can bind with the immobile receptors (pink) on the substrate (grey) reversibly. (b) Illustration of the $\kappa$-$\mu VT$ Monte Carlo simulation, in which some receptors (red) are bound with implicit ligands on the nano-particles (blue) while the others (black) are unbound. In the simulation, the bonds are implicit. {(c) Part of typical snapshots of receptors following various distributions. The global typical snapshots of receptors can be found in SI.}}
\end{figure}

As shown in Fig.~\ref{Fig1}a, we consider that immobile receptors are grafted on a host substrate. 
The nano-particles are controlled by an activity \(z = v_{0} \exp (\beta \mu)\) with $\mu$ the chemical potential of nano-particles and \(\beta=1 / k_B T\), where  $v_0$ is the volume that each particle can explore when bound on the substrate, and $k_B$ and $T$ are the Boltzmann constant and temperature of the system, respectively. Each nano-particle is coated with $\kappa$ mobile ligands, which can bind to the receptors reversibly with the binding free energy $f_B$. {The binding free energy $f_B$ is determined by both the equilibrium constant of ligand-receptor binding in solvent $K_a$ and the configurational entropy penalty due to the constraint of tethering $\Delta S_{\rm conf}$: $\beta f_{B}=-\log K_{a}-k_{B}^{-1} \Delta S_{\rm{conf }}$~\cite{leunissen2010quantitative}.} 
Assuming that the adsorption of each guest particle is independent, we divide the substrate into $N_{\rm max}$ sites, each of which can bind with one guest particle at most. The fraction of sites that are occupied by particles with at least one bond formed is
\begin{equation}\label{eq:theta}
	\theta(z,n_R) =\frac{zq(n_R)}{1+zq(n_R)}.
\end{equation}
Using the unbound site as the reference state, the single-site bound state partition function $q(n_R)$ with $n_R$ receptors can be written as 
\begin{equation}\label{eq:q}
    q\left( n_{R}\right)=\sum_{\lambda=1}^{\min \left(\kappa, n_{R}\right)}Q(\lambda,n_R),
\end{equation}
where $\lambda$ is the number of bonds formed and 
\begin{equation}\label{eqQ}
	    Q(\lambda,n_R)= e^{-\lambda\beta f_{B}} \frac{\kappa ! n_{R} !}{(\kappa-\lambda) ! \lambda !\left(n_{R}-\lambda\right) !}.
\end{equation}

Then the fraction of bound sites or adsorption is 
 \begin{equation} \label{eq:theta_ave}
     \langle \theta \rangle = \left\langle \frac{zq}{1+zq}\right\rangle _{\langle n_R \rangle},
 \end{equation}
where $\langle \cdot \rangle_{\langle n_R \rangle}$ calculates the average over the receptor number distribution with the mathematical estimate $\langle n_R \rangle$. The selectivity parameter is defined as
\begin{equation} \label{eq:alpha}
    \alpha = \frac{{\rm d}\ln\langle \theta \rangle}{{\rm d} \ln \langle n_R \rangle}.
\end{equation}
Since {the higher selectivity} usually appears at small activity~\cite{martinez2011designing,wang2012selectivity,phan2023bimodal}, the time for the substrate to exchange nano-particles with the reservoir to reach equilibrium is very long, which makes the direct Monte Carlo (MC) simulations in 3D systems expensive and inefficient. 
Here we propose a $\kappa$-$\mu VT$ MC simulation method with implicit ligands and bonds in 2D, which enables us to efficiently sample in the additional bond number dimension (see Methods). 
As shown in Fig.~\ref{Fig1}b, we model the multivalent nano-particles as hard disks of diameter $\sigma$ and volume $s_{\rm hd}=\pi \sigma^2/4$ controlled by the chemical potential $\mu$. We assume that one receptor can only bind with the particle covering it, i.e., the center-to-center distance between the receptor and ligand is less than $\sigma/2$. 
The total number of sites $N_{\rm max}=L^2/s_{\rm hd}$, and the average number of receptors per site $\langle n_{R} \rangle=N_R s_{\rm hd} /L^2$. The activity $z =  s_{\rm hd} \exp(\beta \mu) / \Lambda^2$ with $\Lambda$ the de Broglie wavelength. The distribution $p(n_R)$ is numerically sampled by the number of receptors within a 2D spherical window of radius $\sigma/2$. One can see that the $\kappa$-$\mu VT$ MC simulation essentially simulates a monolayer of nano-particles near the host substrate, where nano-particles can bind with receptors on the substrate, and the system exchanges nano-particles with a bulk (3D) reservoir of chemical potential $\mu$ above. The advantage of the $\kappa$-$\mu VT$ model is that one does not need to explicitly simulate the exchange of nano-particles between the host substrate and bulk reservoir through diffusion, which could be very computationally expensive at small activity.

\section*{Results}
\subsection{Receptor uniformity enhances selectivity}
We consider four different types of receptor distributions: anti-hyperuniform (Anti-HU) distributions~\cite{torquato2018hyperuniform}, the Poisson distribution, stealthy hyperuniform (SHU) distributions~\cite{florescu2009designer,torquato2015ensemble} and a square lattice (Fig.~S1). In equilibrium, Anti-HU can describe systems close to a critical point, and SHU describes the disordered systems with long range correlations~\cite{torquato2018hyperuniform}. 
{All those distributions are statistically homogeneous point processes and follow the central limit theorem, i.e., they can be approximated by Gaussian distributions at the large $\langle n_R \rangle$ limit~\cite{torquato2021local}}. 
The spatial uniformity of a receptor distribution at given $\langle n_R \rangle$ can be characterized by the relative local number variance $\sigma^2_{n_R}/\langle n_R \rangle$. For the Poisson distribution in 2D, $\sigma^2_{n_R}/\langle n_R \rangle = 1$. For a perfect square lattice, $\sigma^2_{n_R}/\langle n_R \rangle \sim \langle n_R \rangle^{-1/2}$, which essentially implies that the square lattice is more uniform than the Poisson distribution. 

SHU distributions follow the same scaling with the square lattice. {The configurations are generated by minimizing $\Phi(\mathbf r^N)=\sum_{|\mathbf k|<K} S(\mathbf k)$ using the limited-memory BFGS algorithm~\cite{liu1989limited}, starting from a Poisson configuration with number density $\rho=1$. Here $K=4\sqrt{\pi \chi}$~\cite{torquato2015ensemble} and $\chi=M(K)/[D(N_R-1)]$ denotes the relative fraction of constrained degrees of freedom compared to the total degrees of freedom $D(N_R-1)$ with $M(K)$ the number of independently constrained wave vectors~\cite{torquato2018hyperuniform}.} The prefactor of the distributions depends on the parameter $\chi$ with the larger $\chi$ being more uniform or with smaller density fluctuations. 

On the contrary, in Anti-HU structures, $\sigma^2_{n_R}$ increases faster than $\langle n_R \rangle$, and here we choose configurations that exhibit $\sigma^2_{n_R}/\langle n_R \rangle \sim \langle n_R \rangle^{1/2}$. {The anti-hyperuniform configurations are generated using the algorithm detailed in Ref.~\cite{torquato2021structural}. Specifically, we use the limited-memory BFGS algorithm to minimize $\sum_{|\mathbf k|<K} [\langle S(\mathbf k)\rangle-S_0(\mathbf k)]^2$, in which $K=10$, $\langle S(\mathbf k)\rangle$ is the average structure factor  $S(\mathbf k)=|\sum_{j=1}^{N_R} \exp(-i\mathbf k \cdot \mathbf r_j)|^2/N$ over $N_c=100$ configurations, and the targeted structure factor for various $a$ is $S_0(\mathbf k)=1+ a\exp(-|\mathbf k|)/|\mathbf k|$.} The prefactor of the distributions depends on the parameter $a$ with the larger $a$ being less uniform or with the larger density fluctuations. For each structure, we individually generate 10 snapshots of $10^4$ receptors to sample the spatial distribution and to be used in MC simulations.

\begin{figure}[!b]
\centering
        \includegraphics[width=0.45\textwidth]{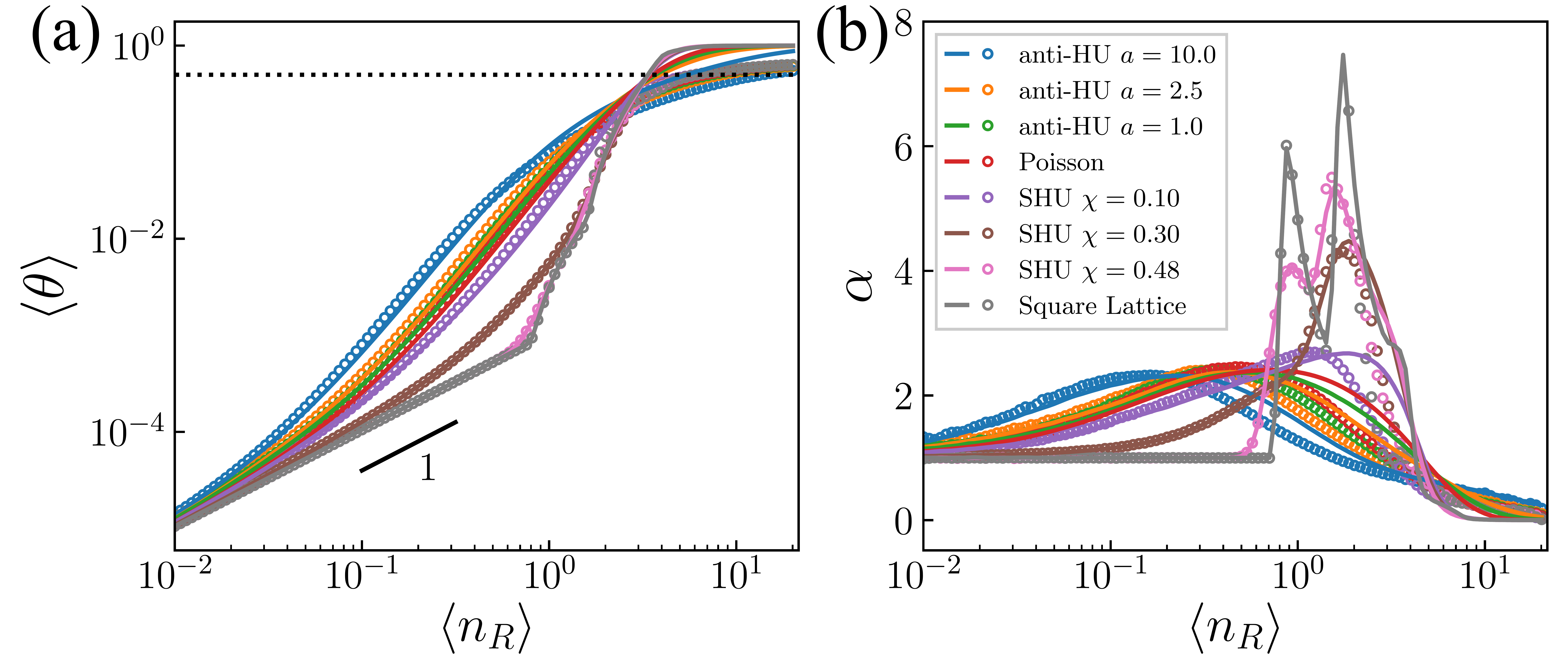}
        \caption{\label{Fig2}{ \textbf{Receptor uniformity enhances selectivity.} Average bound fraction $\langle \theta \rangle$ (a) and selectivity parameter $\alpha$ (b) as a function of $\langle n_R \rangle$ with $\kappa=4$, $\beta f_B=-2$ and $\beta\mu=-10$ for various receptors distributions. The solid curves are the theoretical predictions of  Eqs.~\ref{eq:theta_ave} and \ref{eq:alpha}, and the symbols are obtained from simulations. }}
\end{figure}

\begin{figure*}[!bt]
\centering
        \includegraphics[width=1.\textwidth]{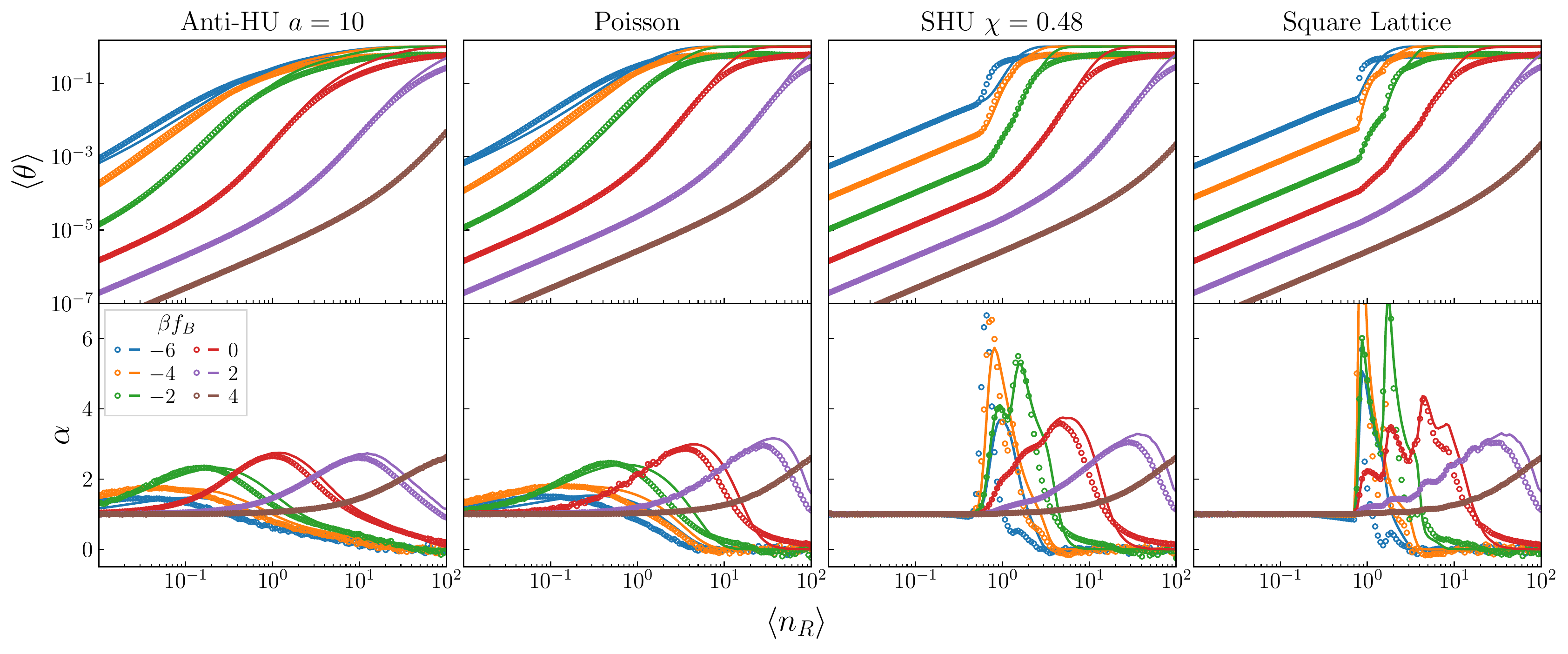}
        \caption{\label{Fig3}{ \textbf{Achieving the highest selectivity by tuning binding free energy.}  $\langle \theta \rangle$ and $\alpha$ as a function of $\langle n_R \rangle$ for various binding free energy $\beta f_B$ for typical receptors structures: Anti-HU $a=10$, Poisson, SHU $\chi=0.48$ and square lattice. The solid curves are the theoretical predictions of Eqs.~\ref{eq:theta_ave} and \ref{eq:alpha}, and the symbols are obtained from simulations. In all simulations, $\kappa=4$ and $\beta\mu=-10$. }}
\end{figure*}
In Fig.~\ref{Fig2}, we plot the average bound fraction $\langle \theta \rangle$ and the selectivity parameter $\alpha$ as functions of $\langle n_R \rangle$ for various receptor distributions.
One can see that Eqs.~\ref{eq:theta_ave} and \ref{eq:alpha} agree quantitatively with computer simulations when $\langle \theta \rangle < 0.5$ (indicated by the dotted horizontal line), and at very large $\langle n_R \rangle$, 
the theoretically predicted $\langle \theta \rangle$ is larger. This discrepancy is due to the fact that the excluded volume effect between nano-particles is not considered in the mean field theory, which overestimates the adsorption at high density. 
When $\langle n_R \rangle$ is small, with increasing $\langle n_R \rangle$, less uniform distributions lead to larger value of $\langle \theta \rangle$. 
{It is because that $\langle \theta \rangle \approx z \langle q \rangle$ according to Eq.~\ref{eq:theta_ave},} where the average bound state partition function over the receptor distribution $\langle q \rangle$ is a convex function (see SI),
hence $\langle q \rangle$ increases with increasing the variance of the distribution $\sigma^2_{n_R}$. 
As shown in Fig.~\ref{Fig2}, with increasing the uniformity, i.e., from Anti-HU to the Poisson, SHU structures and square lattice, the selectivity $\alpha_{\max}$ increases monotonically, and for SHU structures and square lattice, $\alpha_{\max}$ is even larger than $\kappa = 4$. This is intriguing as it has been accepted that weaker binding energy enhances selectivity, of which the upper bound of $\alpha_{\max}$ is $\kappa$~\cite{martinez2011designing}.

\begin{figure*}[!bt]
\centering
        \includegraphics[width=1.\textwidth]{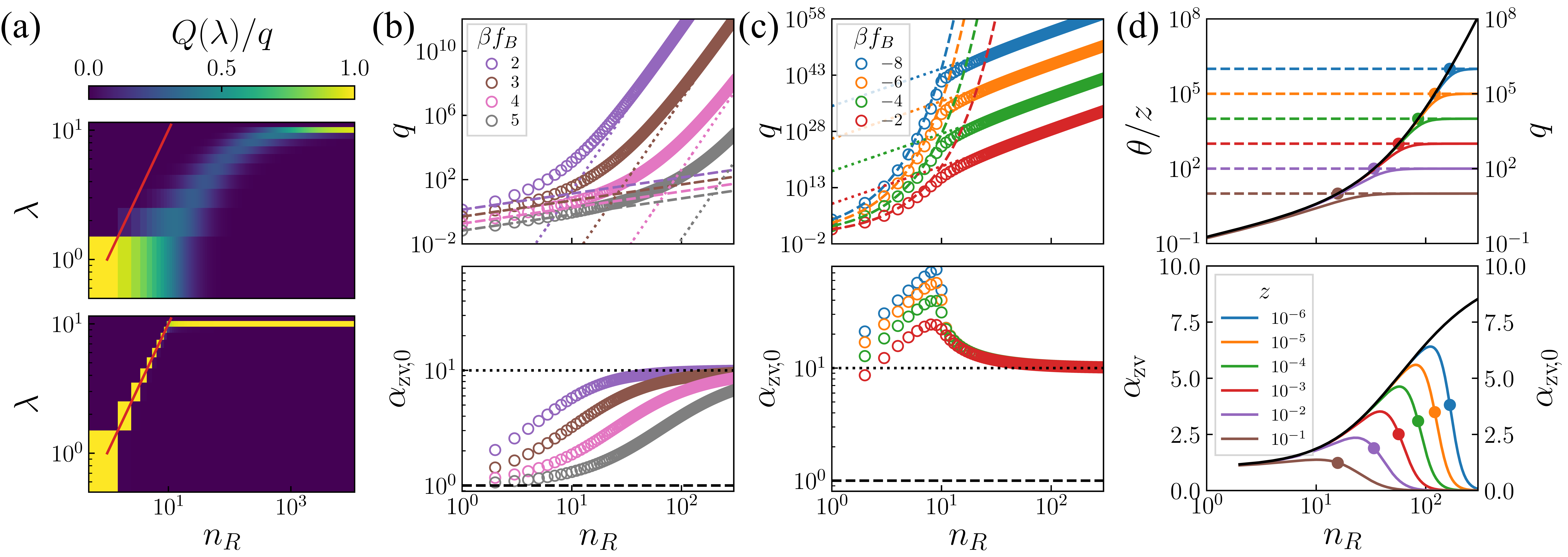}
        \caption{\label{Fig4}{ \textbf{Multivalent binding of nano-particles in the  {zero variance} scenario.}
        (a) The probability for $\lambda$ bonds formed on bound guest nano-particles $Q(\lambda)/q$ as a function of $n_R$ for (upper) $\beta f_B=4$ and (lower) $-4$. 
        (b,c) The {zero variance} bound state partition function $q$ and {zero variance} selectivity at {low activity limit} ${\alpha_{\rm zv,0}}$ as a function of $n_R$. Open symbols are the analytical results from Eq.~\ref{eq:q}. Colored dotted lines are from Eq.~\ref{eq:q_powerlaw}. Colored dashed lines are (b) $q= n_R e^{-\beta f_B}$ and (c) Eq.~\ref{eq:q_exponential}, respectively. Black dashed and dotted lines indicate ${\alpha_{\rm zv,0}}=1$ and $\kappa$, respectively. 
(d)  $\theta/z$ (upper colored solid curves), $q$ (upper black curve) and  ${\alpha_{\rm zv}}$ (lower colored curves), ${\alpha_{\rm zv,0}}$ (lower black curve) as a function of $n_R$ with $\beta f_B = 4$ for various activity $z$. Symbols indicate the position of $n_R^{\rm s}$, which is the crosspoint of  $\theta/z$ and $1/z$ (upper colored dashed curves). In all calculations, $\kappa=10$.}}
\end{figure*}  

\subsection{Achieve the highest selectivity by tuning binding energy}
In Fig.~\ref{Fig3}, we plot $\langle \theta \rangle$ and $\alpha$ as functions of $\langle n_R \rangle$ for binding energy from strong ({$\beta f_B=-6$}) to weak ($\beta f_B=4$) of various receptor distributions. 
For the receptor structures of the Poisson and Anti-HU distributions, $\alpha_{\rm max}$ increases monotonically with increasing $\beta f_B$, namely weaker binding enhances selectivity, while for SHU structures with $\chi=0.48$ and square lattice, $\alpha_{\rm max}$ reaches the maximum at about $\beta f_B = -4$.
To understand this, 
we start with the selectivity parameter ${\alpha_{\rm zv}} = \mathrm{d}\ln\theta / \mathrm{d}\ln n_r$ in the {zero variance} scenario, which is the uniform limit of receptors with $\sigma_{n_R}^2=0$.
As the {superselective} adsorption of nano-particles of interest mostly occurs at low activity, according to Eq.~\ref{eq:theta}, when  $zq\to 0$, $\theta(n_R) \approx zq(n_R)$ and ${\alpha_{\rm zv}} \approx {\mathrm{d} \ln q}/{\mathrm{d} \ln n_R}$.
We define the selectivity parameter {at the low activity limit} as ${\alpha_{\rm zv,0}} = {\mathrm{d} \ln q}/{\mathrm{d} \ln n_R}$~\footnote{Numerically, we calculate the {zero variance} selectivity using the finite difference of $q$ or $\theta$ in the log-log scale: ${\alpha_{\rm zv}}=[\log\theta (n_R+1)-\log\theta (n_R-1)]/[\log(n_R+1)-\log(n_R-1)]$ and ${\alpha_{\rm zv,0}}=[\log q (n_R+1)-\log q (n_R-1)]/[\log(n_R+1)-\log(n_R-1)]$, respectively}. 
We plot the probability of forming $\lambda$ bonds on the guest nano-particle in the bound state in Fig~\ref{Fig4}a. 
For weak binding $\beta f_B=4$ and small $n_R$, one can see $q\left( n_{R}\right) \approx Q(\langle \lambda \rangle_{\rm bound}, n_R) $ with the 
most probable bond number $\langle \lambda \rangle_{\rm bound} \approx 1$ ({yellow region} in upper panel of Fig~\ref{Fig4}a). This implies that there is only one bond formed for the particle bound on the substrate, and the ligands on each particle cannot bind cooperatively.
As shown in Fig.~\ref{Fig4}b, this leads to a linear dependence $q \approx n_R\kappa e^{-\beta f_B}$ and ${\alpha_{\rm zv,0}} \approx 1$ with no superselectivity. With increasing $n_R$,  $\langle \lambda\rangle_{\rm bound}$ approaches $\kappa$ at the large $n_R$ limit due to the restriction from the number of ligands available on nano-particles, and this leads to a power-law dependence of $q$ on $n_{R}$ {(see SI)}:
    \begin{equation}\label{eq:q_powerlaw}
       q \approx Q(\lambda = \kappa, n_R) = e^{-\kappa \beta f_B} \frac{n_R!}{(n_R-\kappa)!} \approx \left( n_R e^{- \beta f_B}\right)^\kappa,
    \end{equation}
and ${\alpha_{\rm zv,0}} \approx \kappa$. Here, $\kappa$ ligands on each particle bind with crowded receptors together, and the emergent combinatorial entropy induces the power-law dependence. 

Differently, at the strong binding $\beta f_B=-4$, although the power-law dependence also appears at the large $n_R$ limit, in the small $n_R$ regime, i.e., $1<n_R<\kappa$, as shown in the lower panel in Fig.~\ref{Fig4}a, the most probable bond number $\langle \lambda\rangle_{\rm bound} \approx n_R$. 
{It is because that the most probable bond number is limited by the number of receptors on each site of the host substrate, rather than the number of ligands on the nanoparticles,} which implies an exponential dependence of $q$ on $n_R$ {(see SI)}:
    \begin{equation}\label{eq:q_exponential}
        q \approx Q(\lambda = n_R, n_R) = e^{-n_R\beta f_B} \frac{\kappa!}{(\kappa-n_R)!} \approx \left( \kappa e^{- \beta f_B}\right)^{n_R},
    \end{equation}
and ${\alpha_{\rm zv,0}} \approx n_R ( \ln \kappa - \beta f_B)\sim n_R$. These are the major results of this work. The derivation of Eqs.~\ref{eq:q_powerlaw} and~\ref{eq:q_exponential} based on the saddle-point approximation method~\cite{angioletti2017exploiting,varilly2012general, angioletti2013communication} can be found in SI.
     As shown in Fig.~\ref{Fig4}c,  ${\alpha_{\rm zv,0}}$ peaks around $n_R=\kappa$, which is independent of the binding energy, and the selectivity ${\alpha_{\rm zv,0}^{\rm max}} \approx \kappa ( \ln \kappa - \beta f_B)$ can be larger than $\kappa$ when $\beta f_B <  \ln \kappa - 1$. Hence, in the {zero variance} binding scenario, $\ln \kappa - 1$ is a threshold of binding energy, lower than which ${\alpha_{\rm zv,0}^{\rm max}}>\kappa$, otherwise  $ {\alpha_{\rm zv,0}^{\rm max}}$ has an upper bound $\kappa$ at $n_R \to +\infty$.    
    
    As shown in Fig.~\ref{Fig4}d, the rescaled bound fractions $\theta/z$  at various activity $z$ collapse at small $n_R$, {while reach reach the plateaus of different height at high $n_R$.} The behavior is qualitatively the same when considering receptor distributions (Fig.~S3).  
This indicates that $z$ does not affect the selectivity parameter when $n_R < n_R^{\rm s}$, where {$q(n_R^{\rm s}) \approx 1/z$} is the bound fraction saturation threshold.    
    Therefore, the {zero variance} selectivity parameter ${\alpha_{\rm zv}} \approx {\alpha_{\rm zv,0}}$ at $n_R$ lower than the threshold, and drops to 0 with further increasing $n_R$. Moreover, when $zq(n_R=1)>1$, i.e., $\beta f_B < \ln (z\kappa)$, the bound fraction saturates even if one nano-particle only binds to one receptor, and no superselectivity occurs. To sum up, an exponential dependence occurs at $\ln (z\kappa) < \beta f_B < \ln \kappa - 1 $ and $1<n_R<\kappa$ in the {zero variance} scenario.

    For the highly uniformly distributed receptors, e.g., SHU $\chi = 0.48$ and the square lattice in Fig.~\ref{Fig3}, the situation is qualitatively the same to the  {zero variance} scenario since the variance of the receptor distribution is small. The largest $\alpha_{\max}$ occurs at an intermediate binding energy with a larger value than $\kappa$. Additionally, the critical $\langle n_R \rangle$ with $\alpha_{\max}$ appears independent of the binding energy {$\beta f_B=-6$ and} $-4$, which is a distinct feature of the exponential dependence. 
    
    For the Poisson and less uniform receptor distributions, i.e., Anti-HU, our numerical evidence shows that $\alpha_{\max}$ increases monotonically with increasing $\beta f_B$ and approaches $\kappa$ at $\beta f_B \to +\infty$ (Fig.~\ref{Fig3}). We believe that this is due to the  relatively large receptor number fluctuations masking the exponential dependence in the intermediate binding energy. 
    As shown in Fig.~S4, in the weak binding limit, receptor uniformity has little effect on $\langle \theta \rangle$, and $\alpha_{\max} \to \kappa$ holds as long as the receptor distribution obeys the central limit theorem and satisfies $\sigma_{n_R}^2< \langle n_R \rangle^2$ at large $\langle n_R \rangle$ (see SI). Therefore, weaker binding enhances the selectivity with an upper bound limit $\kappa$ for those less uniform distributions~\cite{martinez2011designing}. 

\begin{figure}[!t]
   \centering
   \includegraphics[width=0.4\textwidth]{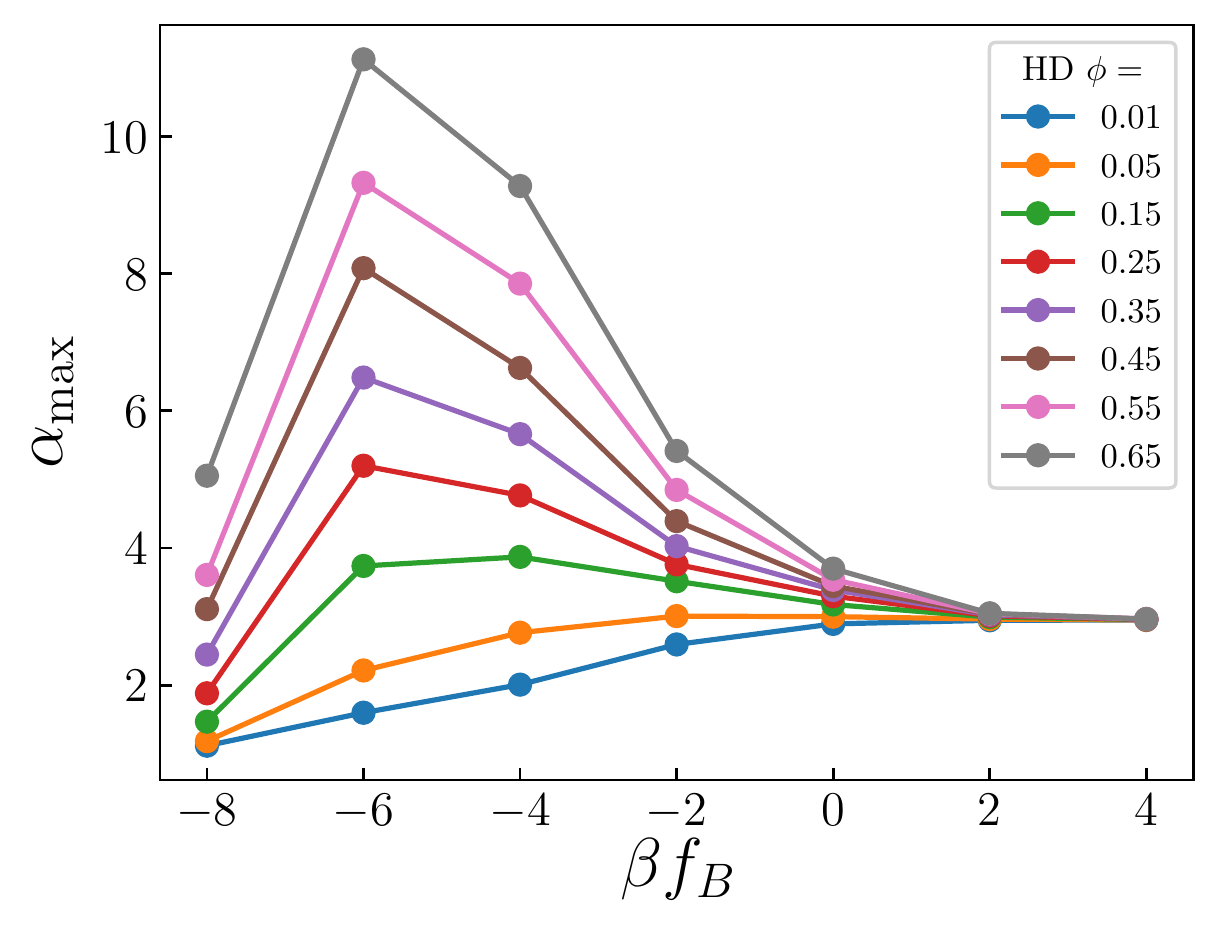}
   \caption{\label{Fig5}{\textbf{Superselectivity of multivalent nano-particle binding on receptors of tunable local uniformity.}  
Selectivity $\alpha_{\max}$ as a function of binding strength $\beta f_B$ for structures obtained from equilibrium hard-disk fluids at various packing fraction $\phi$.   
In all simulations, $\kappa=4$ and $\beta\mu =-10$. }}
\end{figure}

Next we investigate the binding of multivalent nano-particles to receptors with tunable local uniformity. We use the configurations of receptors obtained from equilibrium fluids of hard disks (HD) 
of various packing fraction $\phi$. When $\phi \to 0$, the HD system recovers an ideal gas of the Poisson distribution, i.e, $\sigma_{n_R}^2/\langle n_R\rangle = 1$, and with increasing $\phi$, at certain length scale, $\sigma_{n_R}^2/\langle n_R
\rangle \sim \langle n_R\rangle^{\xi}$ with $\xi<0$, which is the local uniformity induced by the increased short range correlation because of the excluded volume effect (Fig.~S5). The local uniformity of the configuration increases with increasing $\phi$ of HD systems, while at large enough length scale $\sigma_{n_R}^2/\langle n_R\rangle \sim \langle n_R\rangle^{0}$ (Fig.~S5). 
The measured $\alpha_{\max}$ as functions of binding strength $\beta f_B$ for different $\phi$ are shown in Fig.~\ref{Fig5}.
One can see that with small local uniformity, e.g., $\phi = 0.01$, $\alpha_{\max}$ increases monotonically with increasing $\beta f_B$, which is the same as receptors of the Poisson distribution (Fig.~S6). 
However, with increasing $\phi$,  the non-monotonic dependence of $\alpha_{\max}$ on $\beta f_B$ appears when $\phi > 0.15$ (Fig.~S6). This implies that the local uniformity induced by the excluded volume effect can trigger the non-monotonic dependence of selectivity on the binding free energy.

{Here we consider the binding of multivalent nanoparticles on a rigid flat substrate, while on a cell membrane, as more bonds form with the nanoparticles, the membrane roughness decreases. Because the bonds suppress membrane shape fluctuations, which promotes the formation of additional bonds cooperatively~\cite{krobath2009binding,steinkuhler2019membrane}. In our model, we can consider the thermal roughness of flexible substrates by rewriting the partition function of $\lambda$ bonds forming in Eq.~\ref{eqQ} to include an entropy cost, $\Delta S_{\mathrm {mem }}(\lambda)$, that originates from the suppression of membrane shape fluctuations upon $\lambda$ bonds formation. This term depends on the relative roughness of the membranes, $\xi_{\perp} \sim \lambda^{1 / 2}$, and a characteristic length $\xi_{R L}$ that represents the extension of the receptor-ligand complex perpendicular to the membranes~\cite{hu2013binding}. When $\xi_{\perp} \gg \xi_{RL}$, we have $e^{k_B^{-1} \Delta S_{\mathrm{mem}}(\lambda)} \sim \xi_{\perp}^{-1} \sim \lambda^{0.5}$~\cite{krobath2007lateral}. At the large $n_R$ limit, where $q \approx Q\left(\lambda=\kappa, n_R\right)$, the entropy cost term $\Delta S_{\mathrm{mem }}$ is a constant because of the constant number of bonds, and Eq.~\ref{eq:q_powerlaw} remains valid. At this limit, the suppression of membrane shape fluctuation has no qualitative effect on $\alpha_{z v, 0}$. However, an additional term should be added to $\alpha_{z v, 0}$ to account for the contribution of $\Delta S_{\mathrm{mem}}$. This term is given by ${{\rm d} \log e^{k_B^{-1} \Delta S_{\mathrm{mem}} \left(n_R\right)}}/{{\rm d} \log n_R} \approx 0.5$, which suggests that the suppression of membrane shape fluctuations can enhance the superselectivity at this limit.
}


\section{Discussion}
In conclusion, we have investigated the impact of receptor uniformity on the superselective binding of multivalent nano-particles, for which we devised a $\kappa$-$\mu VT$ MC simulation method to compare with analytical theory without any fitting parameter.
We find that receptors that are more uniformly distributed lead to stronger superselective binding of multivalent particles. Specifically, for receptors with SHU structures and square lattice, the selectivity, $\alpha_{\max}$, can be significantly larger than the valence of the nano-particle, $\kappa$, which is the highest level of selectivity that receptors with Poisson and Anti-HU distributions can achieve. Furthermore, for receptors with SHU distributions and square lattice arrangements, the largest $\alpha_{\max}$ occurs at an intermediate strength of binding energy, which is due to the exponential dependence of the bound fraction on the receptor density.
The exponential dependence arises from the restriction of available receptors and is affected by both the binding energy and combinatorial entropy.
 This is different from the binding on receptors with Poisson or Anti-HU distributions, where weaker binding always enhances the selectivity.
These results suggest that for receptors that are highly uniformly distributed, one does not have to use very weak binding energy to achieve high selectivity, and the largest $\alpha_{\max}$ occurs at a certain relatively strong binding that is less affected by non-specific bindings. 

\begin{figure}[!t]
   \centering
   \includegraphics[width=0.45\textwidth]{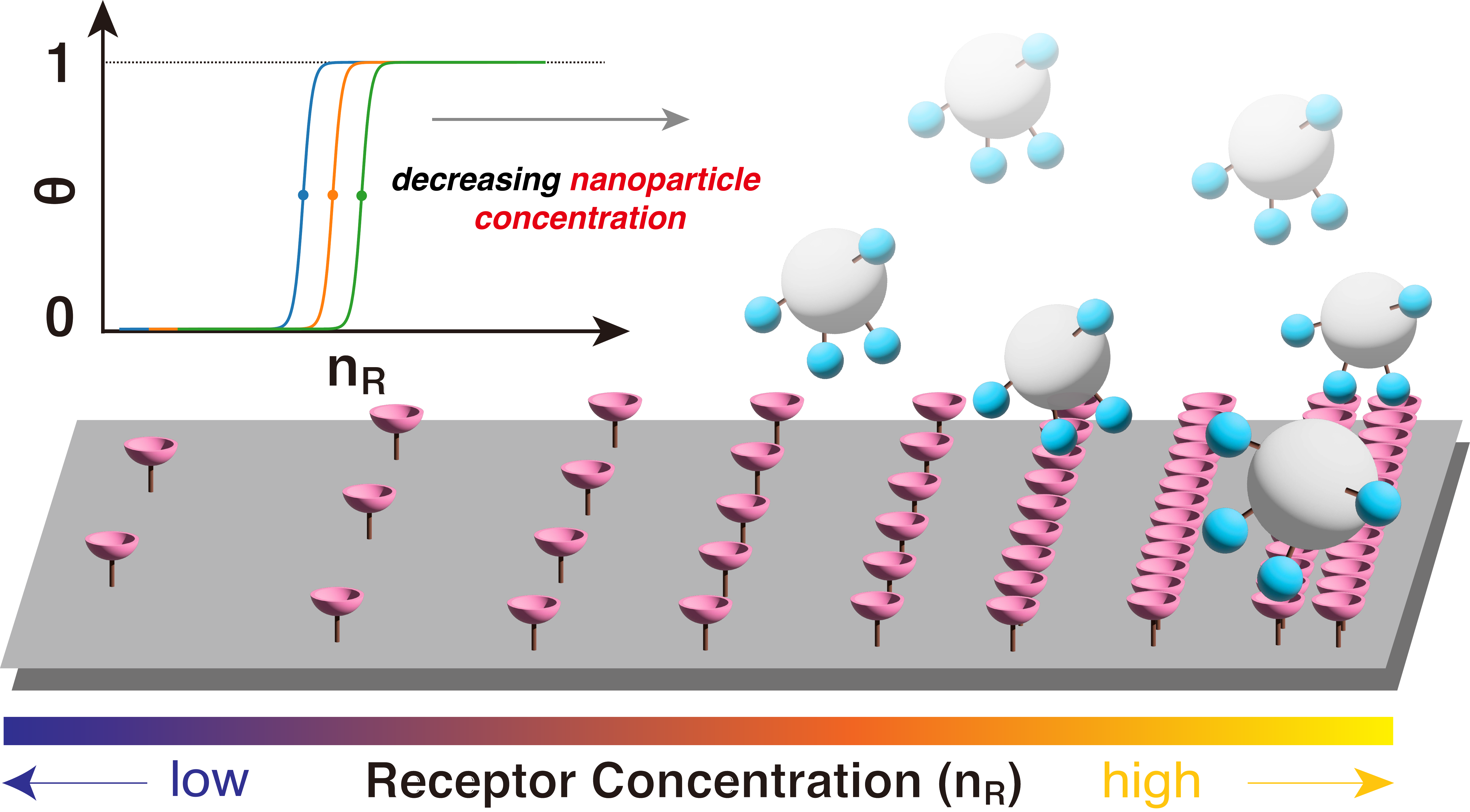}
   \caption{\label{Fig6}{\textbf{A possible design of superselective sensor for multivalent nano-particles.}  
The design consists of arrays of orderly arranged receptors of density $n_R$ increasing from left to right, which can be realized by using DNA origami~\cite{patjacs}. The inset shows typical adsorption curves $ \theta $ of multivalent nano-particles on the bands of receptors of different $n_R$ grafted on the substrate.}}
\end{figure}

Our findings are relevant for designing superselective sensors using multivalent nano-particles, of which a possible design is shown in Fig.~\ref{Fig6}. Based on our results, one can use arrays of orderly arranged receptors of different density using DNA origami to enhance the superselectivity using strongly binding multivalent nano-particles to avoid the influence of non-specific bindings~\cite{patjacs}. {This also suggests the possibility of designing superselective assembly of receptor-patterned colloidal systems~\cite{mognetti2019programmable,song2020programmable}.} 
Additionally, our findings emphasize the significance of receptor distribution in biological systems.
Although in principle, many receptors on cell membranes are mobile, the receptor diffusion time scale on cell membranes, $t_{\rm diff}$, could be much longer than the time it takes to form/break a bond with ligands on nano-particles, $t_{\rm on/off}$. This suggests that they can be considered as effectively immobile receptors.
Furthermore, there are also many immobile receptors on cell membranes that are restricted and compartmentalized due to interactions with the cytoskeleton~\cite{choquet2003role}.
Moreover, we also show that the local uniformity induced by excluded volume effects can trigger the non-monotonic dependence of selectivity on the binding free energy, which suggests that our finding can be expected in systems of relatively densely packed receptors, like the situation in many biological membranes.

\section{Supporting Information Available}
\begin{itemize}
	\item SI.pdf: Simulation methods, additional theory details and figures of the main text.
\end{itemize}

\section{acknowledgments}
We would like to thank Profs. Daan Frenkel and Stefano Angioletti-Uberti for helpful discussions.
This work is supported by the Academic Research Fund from Singapore Ministry of Education Tier 1 Gant (RG59/21) and Tier 2 Grant (MOE2019-T2-2-010).


\section*{References}
\bibliography{ref.bib}

\clearpage
\begin{figure*}
   \centering
   \bf{Graphical TOC Entry}\\
   \includegraphics[width=0.7\textwidth]{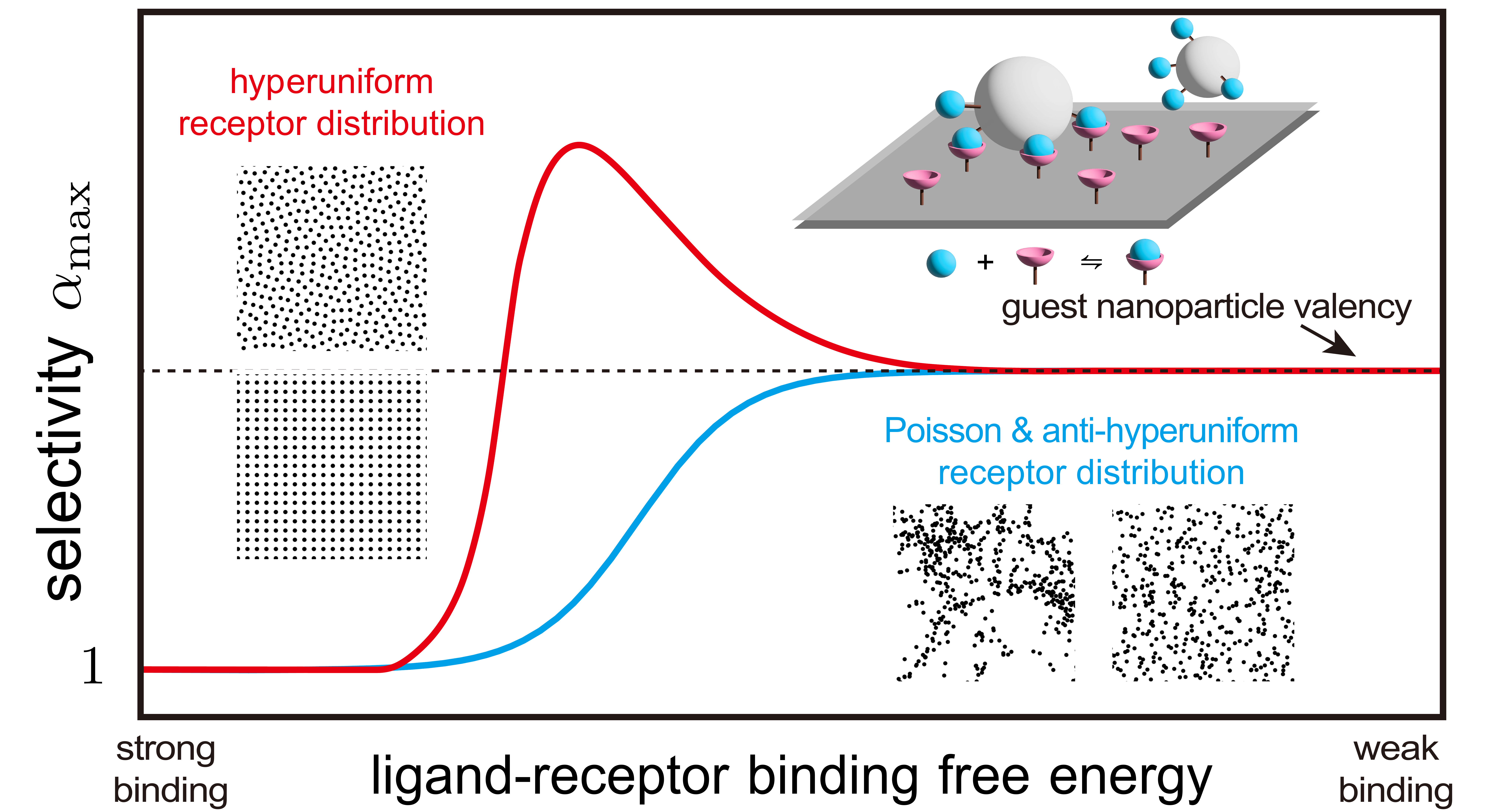}

\end{figure*}

\end{document}